\documentclass[12pt]{article}
\usepackage{latexsym}
\input amssym.def
\input amssym.tex
\font\msym=msbm10

\def\tr{{\mbox{\,tr}}}
\def\Tr{{\mbox{Tr}}}

\def\np#1#2#3{Nucl. Phys. {\bf B#1} (#2) #3}

\def\I_M{{I_{\scriptscriptstyle M\times M}}}

\def\M{{\scriptscriptstyle M}}
\def\N{{\scriptscriptstyle N}}

\def\Com{{\mathop{\hbox{\msym \char  '103}}}}

\def\Z{{\mathop{\hbox{\msym\char '132}}}}

\def\const{{\tilde{\kappa}}}

\begin{document}
\begin{titlepage}
\title{\vskip -60pt
{\small\begin{flushright}
KIAS-P01049\\
hep-th/0202106
\end{flushright}}
\vskip 45pt
Solitons in a Grassmannian $\sigma$-model Coupled to\\ Chern-Simons Term\\}
\vspace{4.0cm}
\author{\\
\\
\\
Jin-Ho Cho,${}^{a}{}^{\natural}$\,\,
Phillial Oh,${}^{a}{}^{\dagger}$ and Jeong-Hyuck Park${}^{b}{}^{\ddagger}$}
\date{}
\maketitle
\begin{center}
\textit{${}^{a}$BK21 Physics Division and Institute of Basic Science\\
Sungkyunkwan University,
Suwon 440-746, Korea} \\
~\\
\textit{${}^{b}$Korea Institute for Advanced Study}\\
\textit{207-43 Cheongryangri-dong Dongdaemun-gu, Seoul 130-012, Korea}
\end{center}
\vspace{1.0cm}
\begin{abstract}
We study a Grassmannian $\sigma$-model coupled to the Chern-Simons term. In the presence of a novel topological term the model admits exact self-dual
vortex solutions which are identical to those of pure Grassmannian model, but the topological charge has a physical meaning as a magnetic flux since the
gauge field is no longer auxiliary. We also extend the theory to a noncommutative plane and analyze the BPS solutions.
\end{abstract}
~\newline
~\newline
~\newline
~\newline
$\overline{\mbox{
E-mail
addresses\,\,:}~{}^{\natural}\mbox{jhcho@taegeug.skku.ac.kr},{}^{\dagger}
\mbox{ploh@dirac.skku.ac.kr},{}^{\ddagger}\mbox{jhp@kias.re.kr}~~~}$
\thispagestyle{empty}
\end{titlepage}
\newpage
\setcounter{footnote}{0}

\section{Introduction}
The $O(3)$ nonlinear $\sigma$-model \cite{bela} and its generalization to CP($N$) and  Grassmannian target space
Gr($N,M$)$\equiv \frac{SU(N)}{SU(N-M)\times U(M)}$ in 2+1 dimensions \cite{zakr} have attracted a great deal of
interest \cite{haldane} especially due to the exact solvability. They admit the exact self-dual BPS soliton
solutions characterized by the second homotopy, $\pi_2(S^2)=\pi_2$(Gr($N,M))=\Z$ \cite{pere}.\\

The fact that the models can be described in a simple manner with the introduction of the auxiliary gauge fields or the composite gauge connections
motivated people to make the gauge fields dynamical either by adding the Maxwell term or the Chern-Simons term \cite{nard,rajeev}.  Especially the
latter has been considered in the context of the fractional spin and statistics \cite{wilz1}. However, gauging the nonlinear $\sigma$-model has been
 focused to $O(3)$ or CP($N-1$)=Gr($N,1$) models up to now. This fact plus the recent upsurge of interest in the solitons of the noncommutative
field theory \cite{harv} motivate us to look at the gauged Grassmannian $\sigma$-model, since the noncommutativity inevitably implies
 the non-Abelian structure in the theory \cite{bak}.\\

In this paper, we consider a Grassmannian $\sigma$-model coupled to the non-Abelian Chern-Simons theory \cite{jack} and show
that  the model is still solvable in the presence of a novel topological term. The model admits exact self-dual
vortex solutions which are identical to those of pure Grassmannian model as in \cite{zakr,macf}. However, the
topological charge has a physical meaning as a magnetic flux since the gauge field is no longer auxiliary. We also
extend the theory to a noncommutative plane and analyze the BPS solutions.\\

The Lagrangian we propose is  with the metric $\eta=\mbox{diag}(-,+,+)$
\begin{eqnarray}\label{lag}
{\cal L}=\,\,\tr\left[-(D_\mu \phi)^\dagger(D^\mu \phi)
-\frac{i}{2}\const \,\epsilon^{\mu\nu\rho}F_{\mu\nu}(\phi^\dagger
D_\rho \phi) -\,\,\left( \phi^\dagger
\phi-\I_M\right)\lambda\right]+{\cal L}_{CS},
\label{lag1}
\end{eqnarray}
where the matter field $\phi$ is an $N\times M$ complex matrix,
$\lambda$ is an $M\times
M$ matrix valued Lagrangian multiplier, and
$\I_M$ is the $M\times M$ identity matrix.
$\const$ is the coupling constant of the newly
introduced interaction term  which is topological being independent of the
metric. The `$\tr$' is the trace over the $U(M)$ local gauge indices.
We take the field $\phi$ to be in
the anti-fundamental representation so that the $U(M)$ gauge transformation is
given with $U\in U(M)$ as
\begin{equation}
\begin{array}{ll}
\phi\rightarrow\phi\,U^{\dagger}\,,~~~&~~~A_{\mu}\rightarrow
UA_{\mu}U^{\dagger}+iU\partial_{\mu}U^{\dagger}\,.
\end{array}
\label{gauge}
\end{equation}
The $U(M)$ covariant derivative is
 $D_\mu\phi=\partial_\mu\phi+i\phi A_\mu$,
and the  field strength is
$F_{\mu\nu}=\partial_{\mu}A_{\nu}-\partial_{\nu}A_{\mu}-i[A_\mu,\,A_\nu]$.
The Chern-Simons term  is
\begin{eqnarray}
{\cal L}_{CS}=\frac{\kappa}{2}\,\epsilon^{\mu\nu\rho}\,\tr\left(A_\mu
\partial_\nu
A_\rho-\frac{2}{3}i A_\mu A_\nu A_\rho \right).
\label{cs}
\end{eqnarray}
Here $\kappa$ is the Chern-Simons coefficient which should be quantized to be
consistent at the quantum level. As for the commutative non-Abelian cases i.e. $M\geq 2$ and also for  any
noncommutative case \cite{0102188},   $2\pi\kappa$ must be an integer. In the next section, we will see that the
aforementioned BPS solutions with the non-zero magnetic flux exist when
$\kappa+\tilde\kappa=0$. The extension to the noncommutative plane will be
done in Sec. 3 and the final section contains some comments and discussions.

\section{BPS Solutions}

Before obtaining the BPS solutions, we first write  the equations of motion
\begin{eqnarray}\label{eom}
&0=\left[-i(D^{\mu}\phi)^{\dagger}\phi+i\phi^{\dagger} D^{\mu}\phi\right]
+\epsilon^{\mu\nu\rho}\left[(\displaystyle{\frac{1}{2}}\kappa
+\const) F_{\nu\rho}-i \const (D_\nu \phi)^\dagger
(D_\rho \phi)\right]\,,&
\label{eqm1}\\
{}\nonumber\\
&0=-D^{\mu}D_{\mu}\phi+\displaystyle{\frac{i}{2}}\const\,\epsilon^{\mu\nu\rho}D_{\mu}\phi
F_{\nu\rho}+\phi\lambda\,,&
\label{eqm2}
\end{eqnarray}
and the Grassmannian constraint coming from the variation of the Lagrangian multiplier
\begin{equation}
\phi^{\dagger}\phi=\I_M\,.
\label{eqm3}
\end{equation}
For the Euclidean plane it is often convenient to introduce the complex
coordinates,
\begin{equation}
z=\frac{1}{\sqrt{2}}(x_{1}+ix_{2})\,,~~~~~\bar{z}=\frac{1}{\sqrt{2}}(x_{1}-ix_{2})\,,
\end{equation}
and hence $\partial=\frac{1}{\sqrt{2}}(\partial_{1}-i\partial_{2}),
\bar{\partial}=\frac{1}{\sqrt{2}}(\partial_{1}+i\partial_{2}),
A_{z}=\frac{1}{\sqrt{2}}(A_{1}-iA_{2}), A_{\bar{z}}=\frac{1}{\sqrt{2}}(A_{1}+iA_{2})$.\newline

In order to obtain the BPS equations one needs to consider the expression of the energy.  The contribution to the
total energy, $E$ comes only from the matter part
\begin{equation}
\begin{array}{ll}
E&\!=\displaystyle{\int}{\rm
d}^{2}x\,\tr\left[(D_{0}\phi)^{\dagger}D_{0}\phi+(D_{z}\phi)^{\dagger}D_{z}
\phi+(D_{\bar{z}}\phi)^{\dagger}D_{\bar{z}}\phi\right]\\
{}&{}\\
{}&\!=\left\{\begin{array}{l}
\displaystyle{\int}{\rm
d}^{2}x\,\tr\left[(D_{0}\phi)^{\dagger}D_{0}\phi+2(D_{\bar{z}}\phi)^{\dagger}
D_{\bar{z}}\phi\right]-2\pi Q\\
{}\\
\displaystyle{\int}{\rm
d}^{2}x\,\tr\left[(D_{0}\phi)^{\dagger}D_{0}\phi+2(D_{z}\phi)^{\dagger}D_{z}\phi\right]+2\pi
Q
\end{array}\right.\\
{}&{}\\
{}&\!\geq |2\pi Q|\,,
\end{array}
\end{equation}
where $2\pi Q$ can be written as a boundary term using the Grassmannian
constraint
\begin{equation}
\begin{array}{ll}
2\pi Q&\equiv i\displaystyle{\int}{\rm
d}^{2}x\,\tr\left[\epsilon^{ij}(D_{i}\phi)^{\dagger}D_{j}\phi\right]
=\displaystyle{\int}{\rm
d}^{2}x\,\tr\left[i\epsilon^{ij}D_{i}(\phi^{\dagger}D_{j}\phi)+F_{12}\right]\\
{}&{}\\
{}&=\displaystyle{\oint_{\infty}}{\rm d}\vec{l}\cdot{\tr
\left(i\phi^{\dagger}\vec{D}\phi+\vec{A}\,\right)}\,.
\end{array}
\label{tr1}
\end{equation}
The saturation of the energy occurs  when the BPS or anti-BPS equations are
satisfied
\begin{equation}
\begin{array}{lll}
0=D_{0}\phi\,,~&0=D_{\bar{z}}\phi&\mbox{:\,BPS~equations}\,,\\
{}&{}&{}\\
0=D_{0}\phi\,,~&0=D_{z}\phi&\mbox{:\,anti-BPS~equations}\,.
\end{array}
\label{BPSeq}
\end{equation}
In either case these BPS or anti-BPS equations determine the gauge field completely.  As  $\phi^{\dagger}D_{\mu}\phi$ is anti-Hermitian  with the
Grassmannian constraint, the BPS or anti-BPS equations imply for all $\mu=0,1,2$
\begin{equation}
\begin{array}{ccc}
0=\phi^{\dagger}D_{\mu}\phi~~~&\Longleftrightarrow&~~~A_{\mu}=i\phi^{\dagger}\partial_{\mu}\phi\,.
\end{array}
\label{Apsi}
\end{equation}
Consequently  $F_{\mu\nu}=iD_{\mu}\phi^{\dagger}D_{\nu}\phi-(\mu\leftrightarrow\nu)$ and $2\pi Q$  in
Eq.(\ref{tr1}) reduces simply to the magnetic flux. \\

For the above BPS or anti-BPS equations to be compatible with the full equations of motion (\ref{eqm1}) and
(\ref{eqm2}),   it is necessary to set $\const+\kappa=0$. With the BPS equations including their implication
(\ref{Apsi}),  the whole equations of motion  reduce  to
\begin{equation}
0=(\kappa+\const)F_{12}\,,
\label{redeqm}
\end{equation}
and $\lambda=\pm F_{12}$ where the upper/lower sign corresponds to the BPS/anti-BPS respectively. We emphasize here that
our BPS solutions are not restricted to the conventional static limit, but can be time dependent. Thus for the BPS states
with nontrivial flux to exist we should set $\const=-\kappa$, and this is the very reason why we introduced the
$\tilde{\kappa}$ term.  Consequently when $\kappa$ is quantized as in the commutative non-Abelian cases or any
noncommutative case, so is  $\const$.  Namely  $\const$  should be an integer divided by $2\pi$   at the quantum level.
It is worthwhile to note that the BPS equations imply the vanishing of the electric field,  $F_{0i}=0$ and hence the BPS
vortices  are purely magnetic. \newline


As the anti-BPS solutions can be obtained simply by $z\leftrightarrow
\bar{z}$, henceforth we focus on solving the BPS equations only.
We first factorize    $\phi=WH$ that is   a product of a  $N\times M$
matrix, $W$  and  a $M\times M$ hermitian matrix, $H$. Then
the Grassmannian constraint~(\ref{eqm3}) determines the Hermitian matrix,  $H^{2}=(W^{\dagger}W)^{-1}$ so that
\begin{equation}
\phi=W\displaystyle{\frac{1}{\sqrt{W^{\dagger}W}}}\,. \label{psiW}
\end{equation}
Substituting this into the BPS equations yields, with $q\equiv\phi\phi^{\dagger}$
\begin{equation}
\begin{array}{ll}
(1-q)\partial_{0}W=0\,,~~&~~(1-q)\bar{\partial}W=0\,.
\end{array}
\end{equation}
Simple solutions for these equations are provided by arbitrary time independent holomorphic matrices
\begin{equation}
W=W_{0}(z)
\end{equation}
on the condition that the inverse of
$W^{\dagger}_{0}(\bar{z})W_{0}(z)$ exists. \newline

More general solutions are available due to the peculiar $GL(M,\Com)$  `symmetry'
\begin{equation}
W\longrightarrow W^{\prime}=W\Lambda\,,
\label{Ltransf}
\end{equation}
where $\Lambda$ is any invertible element in $GL(M,\Com)$.  Under this transformation $H^{-2}$ is covariant, $q$ is invariant and the solution space is
preserved; if $W$ is a solution then so is $W^{\prime}$.
We note that for CP($N-1$) case i.e. $M=1$  the transformation (\ref{Ltransf}) results in   the  ordinary gauge transformation. However for non-Abelian
cases, $M\geq 2$  the transformation can be nontrivial. Only when $\Lambda\in U(M)$ or $\Lambda^{\dagger}=\Lambda^{-1}$ the transformation  reduces to
the  ordinary gauge transformation.  For general $\Lambda$  the transformation is not a symmetry of the action  but generates gauge inequivalent
solutions. Utilizing the `symmetry' we write the BPS solution
\begin{equation}\label{W}
W(t,z,\bar{z})=W_{0}(z)\Lambda(t,z,\bar{z})\,.
\end{equation}

Now we evaluate $2\pi Q$. First, substituting  Eq.(\ref{Apsi}) into the definition of  $Q$ or  Eq.(\ref{tr1})
identifies   $Q$ as the topological number \cite{pere}
\begin{equation}
Q=\frac{i}{2\pi}\displaystyle{\int}{\rm
d}^{2}x\,\epsilon^{ij}\,\tr(q\partial_{i}q\partial_{j}q)\,.
\label{tr2}
\end{equation}
Under the local transformations $\delta q=\delta x^{i}\partial_{i}q$ we get $\delta Q=(3i/{2\pi})\int{\rm
d}^{2}x\,\epsilon^{ij}\tr(\delta q\partial_{i}q\partial_{j}q)=0$, which shows the topological nature of $Q$.  For
the above BPS solutions the corresponding topological number can be written as a surface integral. Straightforward
calculation gives
\begin{equation}
Q=-\frac{1}{2\pi}\displaystyle{\int}{{\rm
d}^{2}x}\,\bar{\partial}\,\tr\left[\displaystyle{\frac{1}
{W^{\dagger}_{0}W_{0}}}W^{\dagger}_{0}\partial
W_{0}\right]= \frac{i}{2\pi}\displaystyle{\oint}{{\rm
d}z}\,\tr\left[\displaystyle{\frac{1}{W^{\dagger}_{0}W_{0}}}W^{\dagger}_{0}\partial
W_{0}\right]\,. \label{tr3}
\end{equation}
To proceed further  we write explicitly
\begin{equation}
\begin{array}{ll}
W_{0}=(w_{1},w_{2},\cdots,w_{\M})\,,~~~~&~~~~w_{a}=v_{a}z^{k_{a}}+{\cal
O}(z^{k_{a}-1})\,,
\end{array}
\label{Ww}
\end{equation}
where $v_{a}$'s are  $N$-component vectors and we can take them to be
orthogonal using the $GL(M,\Com)$ `symmetry'. From $\partial w_{a}=k_{a}w_{a}/z+{\cal O}(z^{k_{a}-2})$ it is
straightforward to get the explicit value of the topological number
\begin{equation} Q=-\displaystyle{\sum_{a=1}^{M}}\,k_{a}\,.
\end{equation}
For the anti-BPS solutions we obtain the positive number,
$Q=\sum_{a}k_{a}$.\newline

Any constant $W_{0}$ corresponds to a vacuum or  a ground state. In fact, the vacuum is a pure gauge as usual.  Since $W^{\dagger}_{0}W_{0}$ can be
diagonalized by a unitary matrix, $U$ with eigenvalues being positive, we can write
\begin{equation}
\begin{array}{ccc}
U^{\dagger}W^{\dagger}_{0}W_{0}U=D~\mbox{:\,diagonal}\,,~~~&~~\hat{\phi}\equiv
W_{0}U\displaystyle{\frac{1}{\sqrt{D}}}\,,~~~&~~
P(z,\bar{z})\equiv\sqrt{D}U^{\dagger}\Lambda(z,\bar{z})\,.
\end{array}
\label{DP}
\end{equation}
Now the vacuum solution reads simply
\begin{equation}
\begin{array}{cc}
\phi=\hat{\phi}\,{\cal U}\,,~~~&~~~A_{\mu}=i\,{\cal U}^{\dagger}\partial_{\mu}{\cal U}\,,
\end{array}
\label{vac}
\end{equation}
where from Eq.(\ref{DP}) the constant $N\times M$ matrix $\hat{\phi}$ satisfies the Grassmannian constraint and ${\cal U}$ is a $M\times M$ unitary
matrix, ${\cal U}=P(P^{\dagger}P)^{-1/2}$. \newline

From the expression for the energy-momentum tensor
\begin{equation}
T^{\mu\nu}=-\left.{\frac{2}{\sqrt{-g}}}\frac{\delta S}{\delta
g_{\mu\nu}}\right|_{g=\eta}=\tr\left(-D^{\mu}\phi^{\dagger}D^{\nu}\phi-D^{\nu}\phi^{\dagger}D^{\mu}\phi
+\eta^{\mu\nu}D^{\lambda}\phi^{\dagger}D_{\lambda}\phi \right)\,,
\end{equation}
for the BPS states  satisfying Eq.(\ref{BPSeq}), $T^{0i}$ and $\epsilon_{ij}T^{0i}x^{j}$ vanish. Hence the BPS
vortices are spinless.

\section{Noncommutative System}
The noncommutative system is described by the same Lagrangian as the commutative one with  all the multiplications
replaced by the Moyal star product. The alternative and equivalent description is the operator formalism where all
the fields are operators acting on a harmonic oscillator type Hilbert space, and  the integration over the
Euclidean plane is replaced by $2\pi\Tr$,  the trace over the Hilbert space.\\

In the previous section, we deliberately organized every expression treating the ordering carefully so that the
noncommutative generalization is to be taken straightforwardly without any reordering of the quantities. Here, we
go over to the noncommutative plane and obtain the operator counterparts - BPS equations, topological number,
magnetic flux as was done in the pure non-commutative CP($N$) case \cite{yang}. Our noncommutative solutions provide one example which has a well defined
commutative limit.\\

The noncommutative plane is defined by a
commutator relation, $[x_{1},x_{2}]=i\theta$ and the Hilbert space is
constructed by the induced  annihilation and creation operators,
$a=z/\sqrt{\theta}$, $\bar{a}=\bar{z}/\sqrt{\theta}$
\begin{equation}
\begin{array}{ll}
|n\rangle=\displaystyle{\frac{1}{\sqrt{n!}}}\,\bar{a}^{n}|0\rangle\,,~~~~&~~~~[a,\bar{a}]=1\,.
\end{array}
\end{equation}
The number, $\sqrt{n\theta\,}$ estimates the radius from the origin in the
noncommutative plane so that $n\rightarrow\infty$ corresponds to the spatial
infinity.  The derivative of a field along the noncommutative coordinate
becomes  $\partial_{i}\phi=[\hat{\partial}_{i},\phi]$ with
\begin{equation}
\hat{\partial}_{i}=\displaystyle{\frac{i}{\theta}}\epsilon_{ij}x_{j}\,.
\end{equation}
In particular, $\partial\phi=[-\bar{z}/\theta,\phi]$ and
$\bar{\partial}\phi=[z/\theta,\phi]$.\newline

In this way, most of  the equations in the previous section can be taken freely for the noncommutative system.
Here we only remark the   subtle issue regarding the trace structure.  Unlike the commutative case the cyclic
property of the trace is not always guaranteed. In order to use the property the quantity in the trace should be
localized or fall off rapidly at spatial infinity. Only in this case one can drop the trace of a commutator or the
boundary term.  From Eqs.(\ref{psiW},\ref{Ww}) the derivative of the matter indeed satisfies this condition.
Therefore  again we can drop the boundary term in Eq.(\ref{tr1}) and justify the cyclic property used  while writing
Eqs.(\ref{tr2},\ref{tr3}).\\

Using  $\Tr [a,\phi]=\lim_{n\rightarrow\infty}\langle n|\,\phi\,a\,|n\rangle$ we recast
the contour integral expression~(\ref{tr3}) for the topological number into the form
\begin{equation}
Q=\Tr\,\tr\left(\left[a,\,(W^{\dagger}_{0}W_{0})^{-1}W^{\dagger}_{0}[\bar{a},\,
W_{0}]\right]\right)
=\displaystyle{\lim_{n\rightarrow\infty}}\,\langle
n|\tr\left((W^{\dagger}_{0}W_{0})^{-1}W^{\dagger}_{0}[\hat{N}\,,
W_{0}]\right)|n\rangle\,,
\end{equation}
where $\hat{N}=\bar{a}a$ is the number operator. From
$[\hat{N},z^{k}]=-kz^{k}$ and Eq.(\ref{Ww}) we have
\begin{equation}
[\hat{N},w_{a}(z)]=-k_{a}w_{a}(z)+{\cal O}(z^{k_{a}-1})\,.
\end{equation}
This  shows the topological number is the same as that of the commutative
case, $Q=-\sum_{a} k_{a}$. Similar analysis also gives the positive number,
$Q=\sum_{a} k_{a}$ for the anti-BPS solution.

\section{Discussions}
Thanks to the novel topological term we introduced,
the Grassmannian model coupled to the Chern-Simons theory
admits exact BPS solutions carrying no electric field. The
topological charge is identified as the quantized magnetic flux, and
this is in contrast to the ordinary
 $O(3)$ nonlinear $\sigma$-model coupled to the Chern-Simons term \cite{nard} where
the gauge potential becomes pure gauge so that the field strength is identically zero \cite{twoform}. Our solution
also contrasts with the exact Jackiw-Pi vortices in the nonrelativistic Chern-Simons model \cite{jackiw}.  They
carry electric charges as well as  the magnetic fluxes which are not quantized. \\

We point out  the  similarity between our BPS vortex solutions and monopoles  in 3+1 dimensional Yang-Mills-Higgs
model. In addition to the exact solvability both objects are electrically neutral and carry the quantized magnetic
flux or charge. Furthermore they are spinless.     This agrees with the close relation between the angular
momentum and charges in $2+1$ dimensions. The angular momentum  of a charge-flux composite is proportional to the
product of the electric charge and the magnetic flux \cite{wilczek}.\\

As the BPS solutions exist only when $\kappa+\tilde\kappa=0$, it would be interesting to see its origin in the
supersymmetrized version of our model, which may dress spins to the BPS vortices as in 3+1 dimensions \cite{ho}.
\\

Finally, we comment that  the solution generating method in the noncommutative theories \cite{harv}  does not work
in our case since it is not compatible with the Grassmannian constraint~(\ref{eqm3}). The method is characterized
by  a nonunitary isometry, $U$
\begin{equation}
\begin{array}{ll}
U^{\dagger}U=I\,,~~~&~~~UU^{\dagger}=(UU^{\dagger})^{2}\neq I\,,
\end{array}
\end{equation}
where $UU^{\dagger}$ is a nontrivial projection operator on the noncommutative  Hilbert space.   Under
the nonunitary isometry  transformation as in Eq.(\ref{gauge}) the equations of   motion transform covariantly but
the Grassmannian constraint is no longer satisfied.
\newpage
~\newline
\newline
\begin{center}
{\large\bf Acknowledgements}
\end{center}
The work of JHC was supported in part by KOSEF through project NO. R01-2000-00021. PO was supported by Korea Research Foundation Grant
(KRF-2001-015-DP0087). The authors acknowledge useful discussions with D. Bak, S. Hong, and B.-H Lee, and especially are indebted to Kimyeong Lee for
valuable comments. The authors also thank the hospitality of the organizers of PIMS-APCTP summer workshop at the Simon Fraser university, during
which this work was initiated.


\begin{thebibliography}{99}
\bibitem{twoform} Nevertheless in \cite{nard}, a nonvanishing gauge invariant two form tensor was constructed as in the 't
Hooft-Polyakov monopole.
\end{thebibliography}
\end{document}